\documentstyle[12pt,aaspp]{article}

\tighten

\begin{document}

\hyphenation{TREE-SPH}

\title
{Simulations of Nonaxisymmetric Instability
in a Rotating Star:  A Comparison Between
Eulerian and Smooth Particle
Hydrodynamics}

\author{Scott C. Smith,{\altaffilmark{1,2}}
Janet L. Houser,{\altaffilmark{2}}
and Joan M. Centrella{\altaffilmark{2}}}

\altaffiltext{1}{Dept. of Physics, Muhlenberg College, Allentown,
PA 18104}
\altaffiltext{2}{Dept. of Physics and Atmospheric Science,
Drexel University, Philadelphia, PA, 19104}

\begin{abstract}
We have carried out 3-D numerical simulations of
the dynamical bar instability in a rotating star
and the resulting gravitational radiation
using both an Eulerian code written in cylindrical
coordinates and a smooth particle
hydrodynamics (SPH) code. The star is modeled
initially as a
polytrope with index $n = 3/2$ and
$T_{\rm rot}/|W| \approx 0.30$, where
$T_{\rm rot}$ is the rotational kinetic energy
and $|W|$ is the gravitational potential energy.
In both codes the gravitational field is purely
Newtonian, and the gravitational radiation is
calculated in the quadrupole approximation.

We have run 3 simulations with the Eulerian
code, varying the number of angular zones and
the treatment of the boundary between the star
and the vacuum.  Using the SPH code we did
7 runs, varying the number of particles, the
artificial viscosity, and the type of initial
model.
We compare the growth rate and rotation speed of
the bar, the mass and angular momentum distributions,
and the gravitational radiation quantities.
We highlight the successes and difficulties of
both methods, and make suggestions for future
improvements.

\end{abstract}


\section{Introduction}

Many of the most interesting astrophysical systems can be
described by the equations of hydrodynamics
coupled to
gravity.  As computers grow more powerful, new numerical techniques
are being developed and computer simulations of astrophysical
systems are gaining importance.  In fact, numerical modeling may be
the only means of getting detailed understanding about certain
phenomena.  Each numerical method has its own
strengths and weaknesses, and the choice of the most suitable method
depends in part on the physical system being studied.  Therefore,
it is important to understand the behavior of the various techniques
in different situations.

One area in which numerical simulations play a key role is the
modeling of astrophysical sources of gravitational radiation.
With the
prospect of several gravitational wave detectors becoming
operational within a decade (e.g.\ \markcite
{Abramovici, et al.\ 1992}; Bradaschia, et al.\ 1990),
the detailed modeling of these sources
has a high priority.
For example, global
rotational instabilites that arise
in collapsing or compact stars
can potentially produce detectable amounts of gravitational
radiation.
A rapidly rotating stellar core that has exhausted its
nuclear fuel and is prevented from collapsing to neutron star size
by centrifugal forces could become unstable and possibly shed
enough angular momentum to allow collapse to a supernova
(\markcite{Thorne 1995}).  Also, a neutron star that is
spun up by accretion
of mass from a binary companion could potentially reach fast
enough rotation rates to go unstable (\markcite{Schutz 1989}).
 Since these sources
are time-dependent, nonlinear, and fully
3-dimensional systems,
calculating the gravitational radiation they produce requires
numerical simulations.

Global rotational instabilities in fluids arise from
nonradial ``toroidal''
modes $e^{\pm im\varphi}$, where
$\varphi$ is the azimuthal coordinate and
$m=2$ is known as the ``bar mode''.
They can be parametrized by
\begin{equation}
	\tau = T_{\rm rot}/|W|,
\label{tau} \end{equation}
where $T_{\rm rot}$ is the rotational
 kinetic energy and $W$ is the
gravitational potential energy
(\markcite{Tassoul 1978};
\markcite{Shapiro \& Teukolsky 1983};
\markcite{Durisen \& Tohline 1985};
\markcite{Schutz 1986}).
 We concentrate on the bar instability
since it is expected to be the fastest growing mode.  This
 instability
can develop by two different physical mechanisms.  The
{\em dynamical} bar
instability is driven by Newtonian hydrodynamics and gravity.
It occurs for fairly large values
of the stability parameter $\tau > \tau_{\rm d}$
and develops on
a timescale of approximately one rotation period. The {\em secular}
instability arises from dissipative processes such as gravitational
radiation reaction and operates in the range
$\tau_{\rm s} < \tau < \tau_{\rm d}$.
 It develops on a timescale of several
rotation periods or longer (\markcite{Schutz 1989}).
The constant density,
incompressible, uniformly rotating Maclaurin spheroids
 have $\tau_{\rm s} \approx 0.14$ and
 $\tau_{\rm d} \approx 0.27$.  For differentially rotating
fluids with a polytropic equation of state,
\begin{equation}
	P  = K \rho^{\Gamma} = K \rho^{1+ 1/n},
\label{poly}  \end{equation}
where $n$ is the polytropic index and $K$ is a constant
that depends on the entropy, early studies indicated that
the secular and dynamical bar
instabilities should occur at about these same values of
$\tau$ (\markcite{Shapiro \& Teukolsky 1983};
\markcite{Durisen \& Tohline 1985}; Managan 1985;
Imamura, Friedman, \& Durisen 1985). Recent work
by Imamura, et al.\ (1995) shows that both
the angular momentum distribution and, to a lesser degree, the
polytropic index affect the value of $\tau$ at which the $m=2$
secular instability sets in.  For the dynamical bar instability
Pickett, Durisen, \& Davis (1996; hereafter PDD) demonstrate
that, for $n=3/2$ polytropes, the $m=2$ dynamical stability limit
$\tau_{\rm d} \approx 0.27$ is valid for initial angular momentum
distributions that are centrally condensed and similar to those
of Maclaurin spheroids.  However, for angular momentum distributions
that produce somewhat extended disk-like regions, both one- and
two-armed spiral instablities appear at considerably lower
values of $\tau$.

As a first step toward understanding realistic sources
 we are simulating the gravitational
radiation emitted when a rapidly rotating star,
modeled initially as a polytrope
with $n = 3/2$ ($\Gamma = 5/3$), becomes
dynamically unstable. Newtonian gravity is used, and
the gravitational radiation produced is calculated in the
quadrupole approximation.  The back reaction of the
gravitational radiation on the star is not included.
We have chosen the case $\tau \approx 0.30$,
which is just above the dynamical stability limit and so might
reasonably approximate a star that spins up (due to collapse or
accretion) and goes unstable.  This case has also been studied
numerically and analyzed using the linearized tensor virial
equations (TVE; see Chandrasekhar 1969)
by Tohline, Durisen, \& McCollough (1985; hereafter TDM),
 so their results are
available for comparison.

We have carried out simulations of the dynamical bar instability
using two very different computer
codes, each based on numerical techniques actively used
in  astrophysics.
One of these is a 3-D Eulerian hydrodynamics code
written in cylindrical coordinates
with monotonic advection.  The other is a
smooth particle hydrodynamics (SPH) code with variable smoothing
lengths and individual particle timesteps.  Since the SPH code
is Lagrangian, gridless, and fully adaptive, it is intrinsically
very different from the Eulerian code.
By running the same calculation on these two
codes, we hope to gain a better understanding of the
relative merits of these methods in modeling the dynamical bar
instability.

An earlier comparison between the results of
using Eulerian and SPH codes to model
the dynamical instability
was carried out by Durisen et al.\ (1986;
hereafter DGTB).
They used rapidly rotating polytropes with
$n=3/2$ and considered the cases
 $\tau \approx 0.33$ and
$\tau \approx 0.38$.
Since they
were studying this instability in the context of star formation,
they did not calculate the gravitational radiation generated.
They
used two different Eulerian codes, one with cylindrical
coordinates (the same one
used by TDM)
and the other with spherical coordinates.  Both of these
used the diffusive donor cell advection and fairly
low resolution.
The SPH code used a
smoothing length that was the same for all particles and
 varied in time,
and a fairly small number of
particles.
Our study takes advantage of more modern and accurate numerical
methods, and focuses on the gravitational radiation generated
by the dynamical instability in compact stars.

This paper is organized as follows.
In \S~\ref{num-tech} we briefly
 describe the numerical techniques used in the two codes,
and in \S~\ref{grav-rad} we discuss the
calculation of gravitational radiation using the
quadrupole approximation.
The initial conditions are presented
in \S~\ref{init-cond}.
The results of modeling the bar instability using the
Eulerian code are given in \S~\ref{evol-bar-Eul},
and the results of using the SPH code in
\S~\ref{evol-bar-SPH}.
We compare the Eulerian and SPH results
in \S~\ref{comparisons} and present our
conclusions in \S~\ref{conclusions}

\section{Numerical Techniques}
\label{num-tech}

Both of the computer codes used in this study solve the equations
of hydrodynamics coupled to Newtonian gravity.  The matter is taken
to be a perfect fluid with equation of state
\begin{equation}
	P=(\Gamma - 1)\rho\epsilon,
\label{gamma-law}  \end{equation}
where $\epsilon$ is the specific internal energy.  Each code has been
subjected to a variety of tests to insure its accuracy and stability.
In this section we present a brief description of each code,
referring the reader to the literature for further details.

\subsection{Eulerian Code}

We use the 3-D Eulerian hydrodynamics code developed
by Smith, Centrella, \& Clancy (1994; see also Smith 1993;
Clancy 1989).  This code is written in
cylindrical coordinates
$(\varpi,z,\varphi)$ with variable spatial zoning.  This
is useful for representing
 rotating configurations, including bars, toroids, and more
complicated geometries, all of which may exhibit substantial
rotational flattening.
The hydrodynamical equations are solved using time
explicit differencing with operator splitting
(Wilson 1979; Bowers \& Wilson 1991).
Although the code has the option of allowing the
grid to move in the $\varpi$ and $z$ directions,
for simplicity we hold both grids fixed
for the models presented in this paper.
 We impose reflection
symmetry through the equatorial plane $z=0$, and calculate
the full range of the angular coordinate $\varphi : 0 - 2\pi$.

In Eulerian hydrodynamics fluid is transported
from one grid zone to another, and it is
important to obtain an accurate value for the quantity
crossing the zone face.
The simplest such advection scheme is the donor cell method,
which is only accurate to first order
and produces large numerical diffusion (Bowers \& Wilson 1991).
To achieve better accuracy and less diffusion, the advection
terms can be updated using an interpolation method that preserves
monotonicity in the quantity being advected.  This code uses
a monotonic advection scheme developed by LeBlanc
(Clancy 1989; Bowers \& Wilson 1991), with all spatial finite
differences in the advection phase being second order.
The spatial differencing in the Lagrangian phase is first
order except for the ``PdV'' term, which is second order.
The code uses first order differencing in time with operator
splitting; in general, this results in a scheme that is
somewhat better than first order in time, but not quite
second order. The method of consistent advection is used
for the angular momentum transport (Norman, Wilson,
 \& Barton 1980; Norman \& Winkler 1986).
Shocks are handled
using a standard artificial viscosity.
  For the bar instability runs
presented in this paper, shocks occur during the later stages
of the evolution, when the spiral arms expand supersonically
and merge.  This shock heating and dissipation generates
entropy.

The Newtonian gravitational potential is calculated by solving
Poisson's equation on the cylindrical grid, with the
boundary conditions at the edge of the grid
specified using a spherical multipole expansion.
 In finite difference
form this becomes a large, sparse, banded matrix equation which we
solve using a preconditioned conjugate gradient method with diagonal
scaling (Press, et al.\ 1992; Meijerink \& Van Der Vorst 1981).
This is a simple and
efficient method that requires a minimum of memory overhead, since
it does not need to store the entire matrix being inverted and
takes advantage of existing arrays already set aside for temporary
storage in the code.  Comparison tests with other sparse matrix
solvers showed that this method produces solutions with the same
accuracy using significantly less CPU time (Smith, Centrella, \&
Clancy 1994).  Such memory and time considerations are very
important for the successful implementation of a fully
3-D Eulerian code.

\subsection{TREESPH}

SPH is a gridless
Lagrangian hydrodynamics method that models the fluid
as a collection of fluid elements of finite extent described
by a smoothing kernel (Lucy 1977; Gingold \& Monaghan 1977;
see Monaghan (1992) for a review).
We have used the implementation of SPH by Hernquist \& Katz
(1989) known as TREESPH.  In this code
each particle is assigned a smoothing length which is
 allowed to vary in both space and time,
thereby achieving roughly the same level of
accuracy in all regions of the fluid. The use of these
variable smoothing lengths as well as individual particle
timesteps makes the program adaptive in both space and time.
TREESPH has the option of evolving either the thermal energy
or a function of the entropy.
 Hernquist (1993) has shown that for adaptive
SPH, in which smoothing lengths vary in time, certain types
of errors do not show up in the total energy if the thermal
energy equation is evolved.  However, if the entropy equation
is used, then conservation of total energy is a good
indicator of the global accuracy of the calculation.
We have chosen to evolve the entropy equation.

The gravitational forces in TREESPH are calculated using a
hierarchical tree method (Barnes \& Hut 1986) optimized for vector
computers (Hernquist 1987).  The particles are first organized
into a nested hierarchy of cells, and the mass multipole moments of
each cell up to a fixed order, usually quadrupole, are calculated. In
computing the gravitational acceleration of a
particle, it is allowed to interact with different levels of the
hierarchy in different ways.  The force due to
neighboring particles is computed by directly summing the two-body
interactions.  The influence of more distant
particles is accounted for
by including the multipole expansions of the cells which
satisfy the accuracy criterion at the location of each particle.  In
general, the number of terms in the multipole expansions is small
compared to the number of particles in the corresponding cells.  This
leads to a significant gain in efficiency, and allows the use of
larger numbers of particles than would be possible with methods that
simply sum over all possible pairs of particles.

As a Lagrangian method
SPH is  attractive
because the computational resources can be concentrated where
the mass is located, rather than spread over a grid that
can be mostly
empty. In addition the numerical algorithms are simpler and,
in general,
considerably easier to implement than standard Eulerian methods.
SPH has been applied to a
variety of astrophysical problems and is gaining in popularity.
However, it is still a relatively
new method, and less is known about its behavior
in various situations than the Eulerian methods, which have
been developed and used by a much larger number of researchers
over the decades.  Comparison studies such as this one
are therefore of considerable interest.

\section{Calculation of Gravitational Radiation}
\label{grav-rad}

We calculate the gravitational radiation produced in these models
using the quadrupole approximation, which is valid for nearly
Newtonian sources (Misner, Thorne, \& Wheeler 1973).
Since the gravitational field in both codes is purely Newtonian,
we calculate only the production of gravitational radiation
and do not include the effects of radiation reaction.
The spacetime
metric can be written
\begin{equation}  g_{\mu \nu} = \eta_{\mu \nu} + h_{\mu \nu},
\label{gmetric}
\end{equation}
where $\mu,\nu = 0,1,2,3$, $\eta_{\mu \nu} = {\rm{diag}}(-1,1,1,1)$ is
the metric of flat spacetime, and $|h_{\mu \nu}| \ll 1$.  The
gravitational waveforms are given by the transverse-traceless (TT)
components of the metric perturbation $h_{ij}$,
\begin{equation}
r\,h_{ij}^{\rm TT} = 2 \, {\skew6\ddot{
{I\mkern-6.8mu\raise0.3ex\hbox{-}}}}_{ij}\,^{\rm TT},
\label{hij-TT}
\end{equation}
where
\begin{equation}
{{{I\mkern-6.8mu\raise0.3ex\hbox{-}}}}_{ij} =
\int\rho \,(x_i x_j -  {\textstyle{\frac{1}{3}}} \delta_{ij} r^2)
 \:d^3 r
\label{Iij}
\end{equation}
is the trace-reduced quadrupole moment of the source.
Note that we use units in which $c = G = 1$ only when
discussing the gravitational radiation quantities.
Here $r^2=x^2 + y^2 + z^2$ is the distance to the source,
spatial indices $i,j=1,2,3$, and a dot indicates a time derivative
$d/dt$.  For an observer located on the axis at $\theta=0, \varphi
=0$ in spherical coordinates centered on the source, the waveforms
for the two polarization states take the simple form
\begin{eqnarray}
h_{+{\rm ,axis}}&=&\frac{1}{r}( {\skew6\ddot{
{I\mkern-6.8mu\raise0.3ex\hbox{-}}}}_{xx}- {\skew6\ddot{
{I\mkern-6.8mu\raise0.3ex\hbox{-}}}}_{yy}),
\label{hplus-axis}\\
h_{\times{\rm ,axis}} &=&\frac{2}{r} {\skew6\ddot{
{I\mkern-6.8mu\raise0.3ex\hbox{-}}}}_{xy}.\label{hcross-axis}
\end{eqnarray}

The gravitational wave luminosity is defined by
\begin{equation}
L = \frac{dE}{dt} = {\frac{1}{5}} \left\langle
{I\mkern-6.8mu\raise0.3ex\hbox{-}}^{(3)}_{ij}
{I\mkern-6.8mu\raise0.3ex\hbox{-}}^{(3)}_{ij}  \right \rangle
\label{lum} \end{equation}
 and
 the angular momentum lost through gravitational radiation is
\begin{equation}
\frac{dJ_i}{dt} = {\frac{2}{5}} \epsilon_{ijk}
 \left\langle {I\mkern-6.8mu\raise0.3ex\hbox{-}}^{(2)}_{jm}
{I\mkern-6.8mu\raise0.3ex\hbox{-}}^{(3)}_{km}  \right \rangle,
\label{dJidt} \end{equation}
where the superscript $(3)$ indicates 3 time derivatives, there is
an implied sum on repeated indices, and the angle-brackets
indicate an average over several wave periods.  For these
burst sources such averaging is not well-defined; therefore we display
the unaveraged quantities ${\textstyle{\frac15}{I\mkern-6.8mu
\raise0.3ex\hbox{-}}^{(3)}_{ij}
{I\mkern-6.8mu\raise0.3ex\hbox{-}}^{(3)}_{ij}}$ and ${\textstyle
{\frac25}\epsilon_{ijk}{I\mkern-6.8mu\raise0.3ex\hbox{-}}^{(2)}_{jm}
{I\mkern-6.8mu\raise0.3ex\hbox{-}}^{(3)}_{km}}$ below.  The energy
emitted as gravitational radiation is
\begin{equation}
\Delta E = \int L\; dt
\label{delta-E} \end{equation}
 and
the angular momentum carried away by the waves is
\begin{equation}
\Delta J_i = \int (dJ_i/dt)dt.
\label{delta-Ji} \end{equation}

The expressions given above for calculating gravitational radiation
in the quadrupole approximation are all functions of at least the
second time derivative of ${I\mkern-6.8mu\raise0.3ex\hbox{-}}_{ij}$.
The standard quadrupole formula consists of using the
definition~(\ref{Iij}) for ${I\mkern-6.8mu\raise0.3ex\hbox{-}}_{ij}$
in these equations.
In an Eulerian code, ${I\mkern-6.8mu\raise0.3ex\hbox{-}}_{ij}$ may
be calculated directly by summing over the grid and the time
derivatives may be taken
numerically.  However, this successive application
of numerical time derivatives can introduce a great deal of noise
into the calculated quantities, especially when the time step varies
from cycle to cycle.

To reduce this problem, Finn \& Evans (1990)
have developed two partially integrated versions of the
standard quadrupole formula that
eliminate one of the time derivatives; they call these
the momentum divergence and the first moment of
momentum formulae.
In these expressions,
${\skew6\dot{{I\mkern-6.8mu\raise0.3ex\hbox{-}}}}_{ij}$
is calculated directly by integrating fluid quantities over
the grid.
Since eliminating one numerical time derivative
in a finite difference code
  greatly increases the
signal-to-noise ratio,
both of these methods significantly reduce the
high frequency numerical noise and produce much cleaner
waveforms than the standard quadrupole formula.

Both of these formulae from  Finn \& Evans (1990)
have been implemented
 in the Eulerian hydrodynamics
code to calculate the gravitational radiation quantities; details
are given in Smith, Centrella, \& Clancy (1994). Since
the resulting waveforms are very similar,
we only show the waveforms obtained using
the first moment of momentum expression in this paper.
Note that this expression gives
${\skew6\dot{{I\mkern-6.8mu\raise0.3ex\hbox{-}}}}_{ij}$;
the waveform requires taking another time derivative
to obtain
${\skew6\ddot{
{I\mkern-6.8mu\raise0.3ex\hbox{-}}}}_{ij}$, and
the luminosity requires still another
derivative.  When these derivatives are
calculated
numerically the signals can still be dominated by
noise, especially when the time step is changing
significantly from cycle to cycle.
This problem was solved by passing
 the data through a filter to smooth
it after
${\skew6\dot{{I\mkern-6.8mu\raise0.3ex\hbox{-}}}}_{ij}$
was calculated, and again after each
numerical derivative was taken.
These techniques produce smooth
profiles for both the waveforms and luminosities, as
shown below.

The gravitational radiation is computed in TREESPH using the
method of Centrella \& McMillan (1993) to calculate
$\skew6\ddot{{I\mkern-6.8mu\raise0.3ex\hbox{-}}}_{ij}$
 analytically from the SPH equations of motion.
With this, the gravitational waveforms are calculated
directly  using the
particle positions, velocities, and accelerations which are
already available in the code.
The resulting waveforms are very smooth functions of time and
require no filtering or smoothing to remove numerical noise.
However, the luminosity and angular momentum lost by
gravitational radiation
do contain the time derivative of the
particle acceleration, which is
taken numerically and therefore introduces
some noise.
We have chosen to smooth the luminosity data
for the SPH runs using simple averaging over a fixed interval
of $0.1 t_{\rm D}$
centered on each point. Here, $t_{\rm D}$ is the dynamical time
defined in equation~(\ref{tD}) below.
 In general this procedure makes a
negligible change in the integrated luminosity~(\ref{delta-E}),
which gives the energy emitted as gravitational radiation, and
produces very smooth profiles (Centrella \& McMillan 1993).
The profiles of the angular momentum lost to gravitational
radiation are not smoothed.

\section{Initial Models}
\label{init-cond}

The initial conditions for our simulations are rotating
axisymmetric equilibrium models having $\tau \approx 0.30$ and
polytropic index $n=3/2$.  The bar instability then grows from
nonaxisymmetric perturbations of these equilibrium spheroids. In
this section, we briefly describe the construction of
 these initial models and
their representations in the Eulerian code and in
TREESPH.

The self-consistent field method of Smith \& Centrella (1992)
is used
 to generate the axisymmetric equilibrium models.  This is based
on the earlier work of Ostriker \& Mark (1968) and Hachisu (1986),
and derives from an integral formulation of the equations of
hydrodynamic equilibrium which automatically incorporates the
boundary conditions.  We use a cylindrical grid $(\varpi,z)$
with uniform zoning.
An initial ``guess'' density distribution is given, and the
gravitational potential is calculated using a Legendre polynomial
expansion to solve Poisson's equation (Hachisu 1986).  A rotation
law of the general form $j(m) = j(m(\varpi))$ is specified,
where $j(m)$ is the specific angular momentum and $m(\varpi)$ is the
mass interior to the cylinder of radius $\varpi$ (Ostriker \&
Mark 1968).  Following the convention of earlier work
(Bodenheimer \& Ostriker 1973; TDM;
 DGTB; Williams \&
Tohline 1987, 1988) we use the rotation law for the uniformly
rotating, constant density Maclaurin spheroids, which can be
written in the dimensionless form
\begin{equation}
	h(m) = \frac{M}{J}j(m)
= \textstyle{\frac{5}{2}}(1 - (1-m)^{2/3}),
\label{rotn-law}  \end{equation}
where $J$ is the total angular momentum and $M$ is the total
mass.  Since polytropes do
not have constant density, this produces differentially rotating
models.  The rotation law is used to calculate a rotational
potential, which is then used with the gravitational potential to
compute an improved density distribution.  This process is
iterated until convergence is achieved.

The freely specifiable quantities
 in this method are the
dimensionless
rotation law $h(m)$, the polytropic index $n$,
 and the axis ratio $R_{\rm p}/R_{\rm eq}$,
 where $R_{\rm p}$ is the polar
radius and $R_{\rm eq}$ is the equatorial radius
of the initial model.
To get a dimensional model, we also specify
the maximum density and the entropy, which is given by the
constant $K$ in the polytropic equation of
state~(\ref{poly}).
Upon convergence to a
solution of the equations of hydrodynamic equilibrium,
 this procedure
gives the density $\rho(\varpi)$,
the angular velocity $\Omega(\varpi)$,
the total mass $M$, the total angular momentum $J$, and
the stability parameter
$\tau$.
  Since $\tau$ is not specified initially, some
experimentation with the axis ratio is generally necessary
to achieve a desired value of $\tau$.  In constructing our model
with $\tau \approx 0.30$ we were guided in our choice of input
parameters by the values given in TDM.
 We found that using $R_{\rm p}/R_{\rm eq} = 0.205$ gives
$\tau = 0.301$; note that this configuration is highly flattened
due to rotation. The central rotation period for this
model is $2.15 t_{\rm D}$ and the rotation period for a point
on the equator is $6.90 t_{\rm D}$, where
\begin{equation}
t_{\rm D} = \left
 ( \frac{R_{\rm eq}^3}{GM} \right ) ^{1/2}
\label{tD}
\end{equation}
is the dynamical time for a {\em sphere}
 of radius $R_{\rm eq}$.
To construct the initial model, we used a
uniform cylindrical grid of
$N_{\varpi} = 65$ radial zones
and $N_z = 23$ axial zones.   The
mass distribution extended out to zone 61
in the $\varpi$ direction
and to zone 19 in the $z$ direction.
 This model required 50 iterations to converge to
a solution with a tolerance of $10^{-10}$
 and used 400 seconds of CPU time on the Cray C90
at the Pittsburgh Supercomputing Center (PSC). The accuracy of
this initial equilibrium model can be measured using the
virial relation.  Let
\begin{equation}
VC = \frac{|2T + W + 3\Pi|}{|W|},
\label{virial}
\end{equation}
where $\Pi = \int P dV$ is the volume integral of the
pressure (Hachisu 1986).  For our initial model,
$VC = 7.4 \times 10^{-4}$.

To evolve this model with the Eulerian code, we first
interpolate it onto the
non-uniform grid used in that code. We
trigger the bar instability by imposing a random perturbation
with amplitude $10^{-3}$ on the density in each grid
zone (cf.\ TDM).
The instability then grows from this relatively low noise level,
with the start of the gravitational wave burst occurring at
$\sim 15 t_{\rm D}$.

We have used two different methods to convert the density
$\rho(\varpi,z)$ and angular velocity $\Omega(\varpi)$
produced by the self-consistent field method into a particle
model to be evolved with TREESPH. Both methods use
equal-mass particles.

The first technique is a simple random or
 ``rejection'' method
(Press, et al.\ 1992; Centrella \& McMillan 1993) that
randomly distributes particles within the probability
distribution $\rho(\varpi,z)$, and then assigns the
appropriate angular velocity.  Since the particles are
accepted into the model randomly, and thus independently
of each other, this method results in both positive and
negative density fluctuations about the target
$\rho(\varpi,z)$.  These fluctuations are relatively large.
They trigger the bar instability,
 with the gravitational
wave burst starting immediately (see model T6 below).

Since the perturbations imposed on the Eulerian
models at the initial time are significantly smaller
than those produced by the random
particle method, we developed a technique for
producing a quieter,
``cold'' initial particle model.
In this method a set of equipotential surfaces is
calculated for the equilibrium model produced by the
self-consistent field method
using the gravitational and rotational potentials.
The mass distribution for this model is specified by
calculating the mass interior to these equipotential
surfaces.  To create a particle representation,
we start by placing particles within the
surface boundary of the star at uniform Cartesian
coordinates. The mass interior to the equipotential
surfaces is then used to determine how to relocate
these particles to produce the desired mass
distribution.
 Particle velocities are assigned
using $\Omega(\varpi)$.
The resulting models have considerably
less noise, with the gravitational wave bursts
beginning at $\sim 10 t_{\rm D}$
(Houser \& Centrella 1995).

\section{Evolution of the Bar Instability: Eulerian Runs}
\label{evol-bar-Eul}

We ran three Eulerian models, labeled E1 -- E3; in all these
models,
we use $N_{\varpi} = 64$ zones in the $\varpi$ direction and
$N_z = 32$ zones in the $z$ direction.
To maximize both resolution and
efficiency we use a finer grid in the region initially occupied
by the matter and a coarser grid outside. The $\varpi$ grid
is uniform up to the zone $j=30$, and the $z$ grid is
uniform up to $k=16$.  The zoning is chosen such that the center
of the radial zone $j=25$
 is at the equatorial radius of the initial model
$R_{\rm eq}$ and the center of the axial
zone $k=9$ is at the polar radius
$R_{\rm p}$.  Outside of this uniformly zoned region,
the zone size increases linearly with zoning ratios
$\Delta \varpi_{j+1}/ \Delta \varpi_j = 1.03$ and
$\Delta z_{k+1} / \Delta z_k = 1.1$  The angular grid is
uniform and covers the range $\varphi: 0 - 2\pi$.
For simplicity, the grids are held fixed throughout
the runs.
The grid boundaries for the Eulerian runs
are set at $\varpi = 3.85 R_{\rm eq}$ and
$z = 1.60R_{\rm eq} = 7.91 R_{\rm p}$.  This
 large amount of initially empty space
is necessary to provide room for the star to expand as the bar
mode grows,
and to specify the boundary conditions for the solution
of Poisson's equation accurately (Smith, Centrella, \& Clancy 1994).
We varied the number of angular zones $N_{\varphi}$ and the vacuum
boundary conditions to test the effects of
these parameters on the bar mode instability.
 The properties of the
Eulerian models are summarized in
Table~\ref{E_models_params}.

To study the development of the bar mode quantitatively, we analyze
the density in a ring of fixed $\varpi$ and $z$ using a complex
Fourier integral
\begin{equation}
C_m(\varpi,z) = \frac{1}{2 \pi}
\int_0 ^{2 \pi} \rho(\varpi,\varphi,z) e^{im\varphi} d\varphi,
\label{Cm}
\end{equation}
where $m = 2$ (TDM).
 The normalized bar mode amplitude is
\begin{equation}
|C| = |C_2|/C_0,
\label{bar-amp}
\end{equation}
where $C_0(\varpi,z) = \overline \rho(\varpi,z)$ is the mean
density in the ring.  The phase angle $\phi_m$ is defined by
\begin{equation}
\phi_m(\varpi,z) = \tan^{-1}{\rm Im}(C_m)/{\rm Re}(C_m).
\label{phase}
\end{equation}
The phase information can be used to describe global
nonaxisymmetric structure propagating in the
$\varphi$-direction.  When such a global mode develops out
of the initial noise we can write
\begin{equation}
\phi_m = \sigma_m t,
\label{eigen}
\end{equation}
where the pattern speed of the $m$th structure is
(cf.\ Williams \& Tohline 1987; PDD)
\begin{equation}
W_m(\varpi,z) = \frac{1}{m} \frac{d\phi}{dt} =
\frac{\sigma_m}{m} .
\label{pattern_speed}
\end{equation}
Thus,
$\sigma_2$ is twice the bar rotation
speed, and the
rotation period of the bar is
$T_{\rm bar} = 4\pi/\sigma_2$.

The bar mode amplitude $|C|$ and phase angle $\phi_2$
have been calculated by TDM using the linearized TVE method.
This technique gives exact results for small oscillations of
uniform density, incompressible ellipsoids such as the
Maclaurin spheriods (Chandrasekhar 1969).  It was adapted
to study rotating compressible fluids by Tassoul \&
Ostriker (1968), and applied to rotating polytropes by
Ostriker \& Bodenheimer (1973).  For compressible fluids,
the TVE method gives only approximate results; see TDM for
a discussion.  Nevertheless, it provides a useful point of
comparison for the numerical simulations.
TDM used the Ostriker--Bodenheimer TVE code to calculate
the bar mode growth rate and eigenfrequency for the
case $\tau = 0.301$.
According to their analysis,
the amplitude $|C|$ should grow exponentially with time as
the instability develops, with  $d\ln
|C|/dt = 0.728 \pm 0.038 t_{\rm D}^{-1}$ (we have converted
from their units).  For the eigenfrequency they find
$\sigma_2 = 1.892 \pm 0.094 t_{\rm D}^{-1}$. The errors
quoted by TDM are of the order $\pm 5\%$ and only account
for the expected inaccuracies in the equilibrium models.

\subsection{Results from our Standard Eulerian Model}
\label{sec-eul-std}

Our standard Eulerian model is E2,
which uses $N_{\varphi} = 64$.
 Density contours showing
the development of the bar instability in this model are
presented in
Figure~\ref{E-contours}.  The growth of the $m = 2$ mode
produces a bar-shaped structure.  This rotating bar
develops a spiral arm pattern as mass is shed from the
ends of the bar. The bar and spiral arms
exert gravitational torques, causing angular momentum to be
transported.  The spiral arms expand supersonically
and merge together, causing shock heating and dissipation
in the disk surrounding the central core.
The system remains highly flattened throughout, and
 evolves toward a {\em nearly} axisymmetric final
state.

The distribution of mass $m(\varpi)$
is shown in Figure~\ref{E2-m} for the initial time
(dot-dashed line), the intermediate time
$t = 21.1 t_{\rm D}$ (dashed line), and the final time
$t = 34 t_{\rm D}$ (solid line).  The
angular momentum $J(\varpi)$ is shown for these same three
times in Figure~\ref{E2-j}.
Note that $J(\varpi)$ is normalized by the total
angular momentum in the system at the time, which is less
than the initial value due to non-conservation;
see Table~\ref{E_models_params}.
 We define the core to be all
matter contained within cylindrical radius
$\varpi = R_{\rm eq}$.  In the
final state, the core has $96\%$ of the mass,
and $86\%$ of the angular momentum;
see Table~\ref{E-Models_data-hydro}.

The growth of the bar mode for model E2
is shown quantitatively in
Figure~\ref{E2-bar}, where $\ln |C|$ is plotted versus time
for the ring $\varpi = 0.362 R_{\rm eq}$ in zone $j=10$ in the
equatorial plane $z=0$.  We also checked the growth of
$\ln |C|$ at several other values of $\varpi$ and found that
the growth rate $d \ln |C|/dt$ is essentially independent
of cylindrical radius within the core. This shows that the
bar mode grows at a well-defined rate (TDM).
  To determine the bar growth rate
we fit a straight line through the data points in
Figure~\ref{E2-bar} in
 the time interval during which the function
$\ln |C|$ is growing linearly with time.
The endpoints of this time interval are chosen ``by eye''.
Then, using the definition that a line segment consists of
at least 10 successive points, the slope is calculated by
linear regression for all possible line segments in this
time interval.  The average of these slopes is used to
determine the growth rate.
 For model E2 we find $d\ln |C|/dt = 0.58 t_{\rm D}^{-1}$.

To determine the eigenfrequency $\sigma_2$, we plot
$\cos \phi_2$ as a function of time, and use a
trigonometric fitting routine to calculate
$\phi_2$.
The function $\cos \phi_2$ is used for simplicity,
since $\phi_2$ itself is multi-valued due to the
$\tan^{-1}$ in equation~(\ref{phase}).  The fit
is performed over the same interval used to calculate
the bar mode growth rate.  As a check on this procedure,
we also use an FFT to calculate the frequency spectrum.
For model E2 we obtain the
eigenfrequency $\sigma_2=1.8 t_{\rm D}^{-1}$,
which gives a bar rotation period
$T_{\rm bar} \sim 7 t_{\rm D}$.
Comparison with Figure~\ref{E2-bar} confirms that the
initial exponential growth of the bar mode takes place
over approximately one bar rotation period.

We calculate the growth of
other Fourier components of the density
 in this same ring using
equation~(\ref{Cm}) and normalizing the resulting
amplitudes by $C_0$.  Figure~\ref{E2-modes} shows the
growth of the amplitudes of
the components (a) $m=1$, (b) $m=3$, and
(c) $m=4$.  In each plot, the amplitude of the bar
mode is shown as a solid line for comparison.
As expected, the growth of the bar mode dominates the initial
stage of the evolution, with the other components
 becoming important at later times.  In particular,
Figure~\ref{E2-modes} (c) shows that the $m=4$ mode
starts growing after the bar mode is well into its
exponential growth regime.  The $m=4$ mode also grows
exponentially, but at a faster rate
$d \ln (|C_4|/|C_0|)/dt = 1.1 t_{\rm D}^{-1}$
than the bar mode. Both modes reach their peak amplitudes
at about the same time, then drop to local minima and
grow again. The eigenfrequency of the $m=4$ mode
is $\sigma_4 = 3.4 t_{\rm D}^{-1}$, giving a pattern
speed for this mode $W_4 = 0.85 t_{\rm D}^{-1}$.
Since the pattern speed of the bar mode is
$W_2 = 0.9 t_{\rm D}^{-1} \sim W_4$,
 this suggests that the
$m=4$ mode is a harmonic of the bar mode, and not
an independent mode.  This agrees with the
expectation that if the $m=4$ mode were an independent
mode, then it would grow at a slower rate than the
bar mode.  See  Williams \& Tohline (1987).

In addition, Figure~\ref{E2-modes} (a) shows that the
$m=1$ disturbance grows to nonlinear amplitude after the
bar mode amplitude has reached its maximum value.  Recent
work by Bonnell (1994; see also Bonnell \& Bate 1994) in the
context of star formation shows similar behavior.  PDD also
find that both $m=1$ and $m=2$ modes arise for certain
initial angular momentum distributions.  These issues should
be investigated more closely in future work.

Figure~\ref{E2-bar} shows that the amplitude of the bar mode
peaks at $t \sim 22 t_{\rm D}$ and then drops to a local
minimum at $t \sim 26 t_{\rm D}$.  It then rises to a local
maximum near $t \sim 30 t_{\rm D}$ and subsequently drops
again.  These features
 can also be seen in the density
contours given in Figure~\ref{E-contours} as follows.
 Focus on the
second highest contour starting in frame (b).  This reaches
a maximum bar-like shape between frames (d) and (e),
and then grows more
axisymmetric until around the time of
 frame (g), which is near the time at which the bar mode
reaches its local minimum. The contour again develops
a bar-like shape before becoming more
axisymmetric.
The behavior of the stability parameter $\tau =
T_{\rm rot}/|W|$ is shown as a function of time
by the solid line in Figure~\ref{tau-vs-t-E2};
for comparison, the dashed line gives
$T_{\rm total}/|W|$.
Note that $\tau$ reaches its minimum value when
the amplitude of the bar mode peaks at $t \sim 22 t_{\rm D}$.
It then rises to a local maximum $\tau \sim 0.27$
near the time $t \sim 26 t_{\rm D}$
 when the bar mode amplitude is at its
local minimum,
and falls as the bar mode amplitude
grows again.  This anti-coincidence of the bar amplitude and
$\tau$ results from the fact that the higher amplitude bar
has a greater moment of inertia, which reduces the rotational
kinetic energy.

The behavior of the gravitational wave quantities is
also strongly linked to the amplitude of the bar mode.
Figure~\ref{E2-waves} shows the gravitational waveforms
(a) $rh_+$ and (b) $rh_{\times}$
 for an observer on the axis at $\theta = 0$,
$\varphi = 0$.
The gravitational waveforms show a
strong initial burst that peaks around the same time that
the bar mode reaches its maximum amplitude,
$t \sim 22 t_{\rm D}$. This is followed by a
weaker secondary burst that
peaks around $t \sim 30 t_{\rm D}$, corresponding to the
secondary maximum of the bar mode amplitude.
Figure~\ref{E2-lum-E} shows (a) the gravitational
wave luminosity L, (b) the energy $\Delta E/M$ emitted as
gravitational waves, (c) the rate $dJ_z/dt$ at which angular
momentum is carried away by the waves, and (d) the angular
momentum $\Delta J_z/J_0$ lost to gravitational radiation.
The luminosity $L$ and $dJ_z/dt$ both
show a primary peak at $t \sim 22 t_{\rm D}$
and a secondary peak at $t \sim 30 t_{\rm D}$, separated
by a local minimum at $t \sim 26 t_{\rm D}$.
Some of the interesting gravitational wave properties are
summarized in Table~\ref{E-Models_data-GW}.

\subsection{Results from
 Eulerian Models with Different Parameter Values}
\label{sec-eul-vary}

Model E1 is the same as E2 except that the resolution in the
$\varphi$ direction is reduced by a factor of two, giving
$N_{\varphi} = 32$ angular zones.  We chose to change
$N_{\varphi}$ because
we are primarily interested in the growth of
nonaxisymmetric modes, and thus the angular resolution is
expected to be an important parameter.
The bar growth rate is
$d \ln |C|/dt =0.53 t_{\rm D}^{-1}$, which is $\sim 9\%$
smaller than in E2.  Since both these models have
reasonably long, well-defined linear growth regions
for $d \ln |C|/dt$ we believe that these differences
are real and not just the result of the method used to
compute them.
The eigenfrequency obtained for model E1
is $\sigma_2 = 1.7 t_{\rm D}^{-1}$

TDM (see also Norman, Wilson, \& Barton 1980;
Williams \& Tohline 1987)
 showed that, for the first-order donor cell advection
scheme, the difference between the true growth rate
and the actual growth rate in the code is given by a
numerical diffusion term.  The size of this diffusive
term is proportional to the size of the angular
zones $\Delta \varphi$. They also showed that,
at least to
first order, the eigenfrequency $\sigma_2$ is not
affected by this numerical diffusion.
 Our code uses a monotonic
advection scheme which is significantly less diffusive
than the donor cell method (Bowers \& Wilson 1991;
Hawley, Smarr, \& Wilson 1984), and in fact the
bar mode growth rates we obtain are larger, and
hence closer to the TVE values, than those
found by Tohline and collaborators.
 Nevertheless, we
expect that increasing the size of $\Delta \varphi$
by decreasing $N_{\varphi}$ will also
lower the growth rate in our code, and this is the
behavior that we find in comparing E1 and E2.
The differences in $\sigma_2$ between E1 and E2
are about a factor
of 2 smaller than the differences in the growth rate.

The bar mode amplitude in run E1 peaks at
$t \sim 22 t_{\rm D}$, then drops off and oscillates
around a lower value for $\sim 10 t_{\rm D}$, and
begins to grow again.
Both the mass $m(\varpi)$ and angular momentum
$J(\varpi)$ distributions in run E1 are more spread out
than in run E2, resulting in a core ($\varpi \le
 R_{\rm eq}$) with a smaller mass and angular momentum.
  See Tables~\ref{E_models_params}
and~\ref{E-Models_data-hydro}.

The gravitational waveforms for E1 show a strong initial
burst with maximum amplitude $\sim 7\%$ smaller than in
E2.  This is
 followed by some additional waves, but they are not cleanly
organized into a secondary burst as in E2.
 The luminosity
$L$ shows both a primary and a secondary peak.
And, although the maximum luminosity in E1 is only
$\sim 5\%$ smaller than in E2,
the ratio of the amplitudes of
the primary and secondary peaks in $L$ is $\sim 20$
for E1 and $\sim 9$ for E2.

The mass density within an Eulerian
 grid zone can
never be zero, since this leads to divisions by zero
in the code.  Therefore,
``vacuum'' regions of the grid actually have a very
small mass
density.  However, if allowed to evolve
 unrestricted, these low
density zones can attain very high velocities
and begin to dominate
the timestep calculations.  To prevent this,
special provisions must be made
to handle these ``vacuum'' regions (R. Bowers,
private communication, 1991). We have chosen
to place the following restrictions on
low density zones.
  For a grid zone in which the density is
below a certain threshold value, the velocity is set to
zero.  The
density threshold we use to limit the velocity for our
standard run E2 (as well as for E1) is
$10^{-7} \rho_{\rm max,i}$, where $\rho_{\rm max,i}$
is the maximum density at the initial time.  Also, if
the density in a zone is $< 10^{-10} \rho_{\rm max,i}$,
the internal energy is set to a fixed value that produces
consistency between equations~(\ref{poly})
and~(\ref{gamma-law}).
Finally, the density itself is set to
$10^{-15} \rho_{\rm max,i}$ if it is less than this
value.
These conditions lead to some loss of
energy and momentum
as matter flows into
these cells.  DGTB
report a similar loss of angular momentum that they
attribute to the zeroing of velocities in low density
zones.

To see how the
vacuum restrictions affect the
evolution of model, we ran
a simulation with relaxed vacuum restrictions.
Model E3 is the same as E2 except that the thresholds
specifying the vacuum conditions are less
restrictive, and the velocity is never set to zero.
For run
E3, the velocity in a zone is set to the sound speed
if it exceeds the sound speed and the density is
 below $10^{-14} \rho_{\rm max,i}$.
The threshold below which the internal energy is
set to a fixed low value
 is $10^{-13} \rho_{\rm max,i}$.  The density
itself is set to $10^{-15} \rho_{\rm max,i}$ if it is
less than this value, as in E2.

Table~\ref{E_models_params} shows that
the simulation time covered by E3 is
$23.9 t_{\rm D}$, which is
considerably less than the $34.0 t_{\rm D}$
covered by E2.
Run E3 was not continued
beyond this point because this would have been too
expensive in terms of CPU time.
 Athough less simulation time
is covered, E3 takes almost
as much computer time as E2, with
 the last $1.5 t_{\rm D}$ of simulation time
 for E3 using 5 CPU hours even with a
relaxed Courant condition to allow $20\%$
larger timesteps.  Because of the high velocities
of the matter in the low density zones,
the timestep
continually
decreases throughout the run, and
we believe that it
would take $\sim 100$ hours of CPU time to
run the remaining $\sim 10 t_{\rm D}$
to complete E3. This demonstrates the need for
restrictions on the vacuum zones.

The less restrictive vacuum conditions contribute to better
energy and angular momentum conservation.
By the end of run E3 the model has lost $1.5\%$ of its
initial energy, and $3.4\%$ of its initial angular
momentum.  At this same time in E2, the model has lost
$2\%$ of its energy and $4.9\%$ of its angular
momentum.

Although E3 was not run long enough to evolve the
model completely, sufficient time has elapsed for
some useful comparisons.  The bar mode
 amplitude has peaked and
spiral structure has developed, but the model
 has not yet settled back
to the nearly axisymmetric final configuration
of E2.  The
gravitational wave amplitudes have also peaked
by this time, although the
initial wave burst is not yet complete.
 We can thus compare the
bar mode properties and the peak gravitational wave signals.
  To provide a reasonable
comparison of other properties, we evaluate them for
 E2 at $23.9 t_{\rm D}$ (these entries in
Tables~\ref{E_models_params}
--~\ref{E-Models_data-GW} are labeled
${\rm E}^{\prime}$) and compare
them to the final results for E3.

Examining the bar mode properties of E3 shows that
they are very similar
to those for E2.  The growth rate for E3 is
only slightly smaller, and the
eigenfrequencies are the same to within the
limit of our measurement accuracy.
This is not surprising, since the
growth of the bar mode
 occurs before there is
significant expansion of the model. Also, these
 properties are measured within
the central bulk of the configuration, so they
should not be strongly
affected by the treatment of the vacuum regions.
  An examination of the
density contours, however, shows that the system
expands significantly more when the vacuum
 conditions are relaxed. Figure~\ref{contours-E2-E3}
shows density contours for (a) E2 and (b) E3 at
time $t = 23.9 t_{\rm D}$.  The inner two contours
are nearly identical, but the spiral arms
ejected by the spinning
star extend out to a larger radius than in E2.

The gravitational wave
 pulses are qualitatively
similar but model E3 shows
somewhat reduced amplitudes,
with the maximum amplitude $\sim 9\%$ smaller
than that of E2.
 The difference is more pronounced when
we examine the luminosity, with the
peak  luminosity of
 model E3
$\sim 19\%$ smaller than that of E2.
The peak gravitational radiation amplitudes are
thus somewhat sensitive to the treatment of the
boundary between the fluid and the vacuum,
which affects the outer, lower density regions
of the star.  However, this is not expected to
be a major factor in situations that are not
dominated by expansion, such as rotating
stellar core collapse.

After this work was completed, we learned that
R. Durisen and collaborators have carried out
similar calculations using an Eulerian code (PDD).
They were able to avoid problems arising from the
time step becoming too small without inhibiting
expansion by setting the background density to
a value between $10^{-10}$ and $10^{-7}$
$\rho_{\rm max,i}$ and limiting all velocities in the
background to less than twice the maximum initial
sound speed (R. Durisen, private communication).
We plan to incorporate their suggestions into our
future simulations.

An alternative means of achieving a better treatment of the
interface between the star and the vacuum might be to use the
piecewise parabolic method (PPM; Colella \& Woodward
1984; Davies, et al. 1993), which is known to be
very good at handling discontinuities in the flow.
It also has a higher resolution for a fixed number of
zones than the finite difference scheme used here.
Of course, PPM is also more expensive in terms of
CPU usage.  A comparison of this model run on a PPM
code would be very interesting.

We can also compare our results with other numerical
calculations.
Tohline and collaborators
used a 3-D Eulerian code written in
 cylindrical coordinates.
  Their code does not
solve an energy equation, and therefore has no way to handle
self-consistently the shocks that form as the spiral
 arms expand. Instead, they required that the fluid
maintain the same polytropic equation of
state~(\ref{poly}), and hence the same entropy,
throughout the evolution. They
used donor cell advection, which is known to be very diffusive.
 And, their code assumes ``$\pi$-symmetry'', which
means that the flow is calculated only in the angular
 range $0 \leq
\varphi < \pi$ so that only the even mode
distortions are modeled.
TDM used $N_{\varpi} = 31$, $N_z = 15$ and
$N_{\varphi} = 32$,
with the model extending out to zone 24
 in the $\varpi$-direction
and zone 9 in the $z$-direction; this is essentially the
same as our resolution for E2 and E3.
They found a bar mode growth rate of
$d\ln|C|/dt = 0.22 t_{\rm D}^{-1}$ and an eigenfrequency
 $\sigma_2 = 2.1 t_{\rm D}^{-1}$.  They demonstrated that the
substantial deviation from the TVE result
 for the growth rate is
due to the large numerical diffusion
 in their code; see TDM for
details. Williams and Tohline (1987) modeled the
initial development of the bar instability
in a polytrope with $\tau = 0.31$ using the
same code with $N_{\varpi} = 32$, $N_z = 32$, and
$N_{\varphi} = 64$, again employing $\pi$-symmetry.
This doubling of the number of angular zones increased
the bar mode growth rate to
$d\ln|C|/dt = 0.49 t_{\rm D}^{-1}$; the eigenfrequency
was  $\sigma_2 = 1.9 t_{\rm D}^{-1}$.
 They found that the $m=4$ mode
starts growing exponentially after the bar mode does,
and that it grows at a faster rate. Also, the $m=4$
pattern moves together with the $m=2$ pattern, with
the maxima locked in phase, implying that
the $m=4$ pattern is a harmonic of $m=2$,
and not a distinct mode.  Finally, PDD used a modified
version of Tohline's code which is second order in all
spatial differences, including advection terms, and second
order in time. They calculated the development of the bar
instability in a polytrope with $\tau = 0.304$ using
$N_{\varpi}$ = 64, $N_{z}$ = 16, and $N_{\varphi}$ = 64
{\em without} $\pi$-symmetry.  They obtained
$d\ln|C|/dt = 0.58 t_{\rm D}^{-1}$ and
$\sigma_2 = 2.1 t_{\rm D}^{-1}$ (PDD).

\section{Evolution of the Bar Instability: SPH Runs}
\label{evol-bar-SPH}

We ran a series of 7 models using TREESPH, labeled
T1 -- T7.  All of these models use equal-mass
particles.  The initial state for each model was
produced using the ``cold'' method described
in \S~\ref{init-cond} except for T6, in which the
random method was used.  We vary the number of
particles $N$ and the linear and quadratic
artificial viscosity coefficients $\alpha$ and
$\beta$, respectively.  In all cases, the number of
neighbors that contribute to the smoothing kernel
is chosen to be ${\cal N}_{\rm S} = 64$.
 The properties of the
SPH models are summarized in
Table~\ref{T_models_params}.

\subsection{Results from our Standard SPH Model}
\label{sec-sph-std}

Our standard SPH model is T7, which has
$N=32,914$ particles, $\alpha = 0.25$, and
$\beta = 1.0$.  Figure~\ref{T7-dot-plots}
shows all the particles in this
model projected onto the equatorial plane.  The
system rotates in the counterclockwise direction.
We see the development of the bar and then the
spiral arm pattern as mass is shed from the ends of
the rotating bar.  The gravitational torques exerted
by the bar and spiral arms cause angular momentum to
be transported.  The spiral arms expand
supersonically and merge, causing shock heating.
The system then evolves toward a {\em nearly}
axisymmetric final state.  Throughout its evolution
the system remains flattened.
Figure~\ref{T7-meridional} shows the particle
positions projected onto the $x - z$ plane at the
final time $t = 35 t_{\rm D}$. Note that the disk
around the central core is somewhat
puffed up due to shock heating during the expansion of
the spiral arms.

Density contours in the equatorial plane
for model T7 are shown in
Figure~\ref{T7-contours}, with the frames
corresponding to the same ones displayed in
Figure~\ref{T7-dot-plots}.  To produce these
contours we first use kernel estimation to interpolate
the density of T7 onto the cylindrical grid used for
the Eulerian model E2.  We then calculate contours for the
matter located in the equatorial plane.  The contour levels
are the same as those used in
Figure~\ref{E-contours} for model E2.

The mass distribution $m(\varpi)$ is shown in
Figure~\ref{T7-m} for the initial time (dot-dashed line),
 the intermediate time
$t = 18.2 t_{\rm D}$ (dashed line), and the final time
$t = 35 t_{\rm D}$ (solid line) for model T7.
Figure~\ref{T7-j} shows the angular momentum distribution
$J(\varpi)$ for these same three times.
 As before, we
define the core to be all matter contained within
cylindrical radius $\varpi = R_{\rm eq}$.  In the
final state the core has $90 \%$ of the mass and
$72 \%$ of the angular momentum;
 see Table~\ref{T-Models_data-hydro}.

To study the development of the Fourier components
of the density, we first
interpolate the particle model onto the cylindrical
grid every $0.1 t_{\rm D}$.  We
then use the same procedure developed for the Eulerian
models and analyze the density in the ring
$\varpi = 0.362 R_{\rm eq}$ in zone $j = 10$ in the
equatorial plane using equations~(\ref{Cm})
--~(\ref{pattern_speed}).
  The growth of the bar mode is shown
quantitatively in Figure~\ref{T7-bar}, where
$\ln |C|$ is plotted versus time.
Comparison of Figure~\ref{T7-bar} with
Figure~\ref{E2-bar} shows that the region in which
$\ln |C|$ grows linearly with time is shorter and less
clearly defined in run T7 than in run E2.  Since the
endpoints of this linear region are chosen ``by eye,''
there is a somewhat greater element of subjectivity
in the resulting bar mode growth rate for T7; this
applies to all the TREESPH runs.
For T7 we find that the
growth rate is $d \ln|C|/dt = 0.51 t_{\rm D}^{-1}$
and the eigenfrequency is
$\sigma_2 = 1.9 t_{\rm D}^{-1}$.

Figure~\ref{T7-modes}
shows the growth of the amplitudes of the
components (a) $m = 1$,
(b) $m = 3$, and (c) $m = 4$.
 In each plot, the amplitude of the bar
mode is shown as a solid line for comparison.
As we saw in \S~\ref{evol-bar-Eul}.
 the growth of the bar mode dominates the initial
stage of the evolution.  The $m=1$ and $m=3$
components do not exhibit significant growth.
The $m=4$ mode
starts growing after the bar mode is well into its
exponential growth regime, and grows
at a faster exponential rate
$d \ln (|C_4|/|C_0|)/dt = 1.1 t_{\rm D}^{-1}$.
 Both the $m=2$ and $m=4$ modes reach their peak amplitudes
at about the same time, then drop to local minima and
grow again. The eigenfrequency of the $m=4$ mode
is $\sigma_4 = 3.8 t_{\rm D}^{-1}$, which gives a pattern
speed for this mode $W_4 = 0.95 t_{\rm D}^{-1}$.
Since the pattern speed of the bar mode is
$W_2 = 0.95 t_{\rm D}^{-1} = W_4$,
 this implies that the
$m=4$ mode is a harmonic of the bar mode, and not
an independent mode.

Figure~\ref{T7-bar} shows that the amplitude of the bar
mode peaks at $t \sim 18.5 t_{\rm D}$ and then drops to
a local minimum at $t \sim 25 t_{\rm D}$.  The amplitude
then begins to rise sharply again and levels off around
$t \sim 30 t_{\rm D}$. Compare this behavior with that
seen in the contour plots in Figure~\ref{T7-contours}
by focussing on the innermost contour starting in
frame (b).  This shows the pronounced initial
development of the bar, with the maximum elongation
of the contour occurring around the time of frame (d),
in agreement with Figure~\ref{T7-bar}.
This contour then becomes more axisymmetric until around
the time of frame (g), after which it elongates again
and then grows more axisymmetric.
Thus, in run T7, this contour becomes nearly axisymmetric
before the bar mode amplitude reaches its local minimum.
For comparison,
the behavior of the stability parameter $\tau$
 is shown as a function of time
by the solid line in Figure~\ref{tau-vs-t-T7};
the dashed line gives
$T_{\rm total}/|W|$.
 Note that $\tau$ reaches a local minimum when
the amplitude of the bar mode peaks at $t \sim 18 t_{\rm D}$.
It then rises to a local maximum $\tau \sim 0.27$
 at $t\sim 22 t_{\rm D}$ and drops off again.

The gravitational radiation quantities also exhibit
features corresponding to the behavior of these modes.
Figure~\ref{T7-waves} shows the gravitational waveforms
(a) $rh_+$ and (b) $rh_{\times}$
 for an observer on the axis at $\theta = 0$,
$\varphi = 0$; cf.\ Table~\ref{T-Models_data-GW}.
The gravitational waveforms show an initial burst that
peaks at $t \sim 18 t_{\rm D}$, corresponding to the initial
growth of the bar instability.  After the peak, the amplitude
drops off until $t \sim 22 t_{\rm D}$
and then stays at a nearly constant value for
about one bar rotation period, during which time
the bar mode amplitude reaches its local minimum.
Recall that $T_{\rm bar} = 2 T_{\rm GW}$,
where $T_{\rm GW}$ is the period of the gravitational waves.
The wave amplitude then drops again at
$t \sim 30 t_{\rm D}$ and stays at a nearly
constant amplitude for about another bar rotation period
before the run ends.  These low amplitude waves are produced
by the rotating, slightly non-axisymmetric core in the
final state; cf.\ Figure~\ref{T7-contours}.
Figure~\ref{T7-lum-E} shows (a) the gravitational
wave luminosity L, (b) the energy $\Delta E/M$ emitted as
gravitational waves, (c) the rate $dJ_z/dt$ at which angular
momentum is carried away by the waves, and (d) the angular
momentum $\Delta J_z/J_0$ lost to gravitational radiation.
The luminosity $L$ shows a broad primary peak
centered around
$t \sim 18 t_{\rm D}$, followed by a drop
to a local minumum around $t \sim 23 t_{\rm D}$ and then a
secondary feature at $t \sim 25 t_{\rm D}$.  The signal then
drops again to a nearly constant level for
 $t \gtrsim 30 t_{\rm D}$.

\subsection{Results from SPH Models with Different Parameter
Values}
\label{sec-sph-vary}

To understand how changing the resolution affects the behavior
of the model, compare runs T1, T2, and T3 with
the standard run T7.  These runs differ only in the total
number of particles $N$; see Table~\ref{T_models_params}.
In all these cases, the bar and spiral arms develop as in T7.
Plots of the particles projected onto
the equatorial plane appear visually similar, except that
the extent of the spiral arm pattern increases somewhat
as $N$ increases due to the larger number of particles
available to resolve the outer regions.
Table~\ref{T-Models_data-hydro} shows that the properties
of the cores of these runs are all similar.
The behavior of the Fourier components of the density
 in these runs is also similar,
except that the results for T1 are much noisier due to its
very low resolution.  Although T1
 has the largest bar mode growth
rate,  we attribute this to the lack of a clearly
defined linear growth region for $d\ln |C|/dt$ and therefore
do not consider it to be a reliable indicator of the accuracy
of this model.

The gravitational wave quantities show definite trends with
particle number $N$. As $N$ decreases, the amplitude of the burst
goes down and the structure of the waveforms after the burst
becomes less distinct.  The peak amplitude of the luminosity $L$
also decreases, and the secondary peak becomes a plateau and
then disappears into noise for the poorly resolved run T1.
For the sequence of models T7, T3, and T2, each run has
roughly half
the number of particles as the previous one.
Quantitatively, the amplitude of the burst goes down by
$\sim 10\%$ between model T7 and T3, and another $\sim 10\%$
between T3 and T2.  The peak values of $L$ and $dJ_z/dt$
decrease by $\sim 30\%$ between T7 and T3, and by another
$\sim 20\%$ between T3 and T2.
See Table~\ref{T-Models_data-GW}.

Thus, the larger the number of particles $N$,
 the larger the amplitudes of the gravitational
wave signals.
 It is clear from
Table~\ref{T-Models_data-GW} that the values of these
amplitudes have not yet converged.
One possible means of achieving convergence is simply
to increase $N$.  However, doubling the number of particles
increases the effective grid resolution of the model by
only $\sim 2^{1/3}$, since we are working in 3-D.
Perhaps a better method would be to use non-equal-mass
particles, with the lower mass particles distributed in
the lower density regions and thus increasing
the resolution in the outer parts of the
star (e.g. Monaghan \& Lattanzio 1985).

In numerical simulations, artificial viscosity terms
are typically added to the momentum and thermal
energy equations to provide a dissipative
mechanism that converts the energy jump across the
shock into heat (Bowers \& Wilson 1991).  This smooths
out the discontinuities that occur in shock fronts
while satisfying the Rankine-Hugoniot relations, which
specify the conservation of mass, momentum, and energy
across the shock (Landau \& Lifshitz 1959).  If there
is  no artificial viscosity, the kinetic energy of the
matter passing through the shock is not correctly
converted into heat and there can be large
post-shock oscillations in the fluid.

The standard SPH artificial viscosity contains two
terms, one that is linear in the particle velocity
differences, and another that is quadratic
(Monaghan 1992).  The linear term has
user-specified coefficient
$\alpha$ and tends to dominate for
low Mach numbers, while the quadratic term has
coefficient $\beta$ and is important for higher
Mach numbers.  Typically, the values $\alpha \sim 1$
and $\beta \sim 2$ are used and the shock front is
spread over $\sim 3 - 4$ particle smoothing
lengths (Hernquist \& Katz 1989).
This type of SPH viscosity can introduce shear into
the flow, particularly
through the linear term
(Hernquist \& Katz 1989; Monaghan 1992).  Since shear
can affect the bar mode instability we want to keep
the artificial viscosity coefficients, particularly
$\alpha$, as small as possible while still maintaining
accuracy in the presence of the shock waves that occur
as the ends of the bar and the spiral arms expand
supersonically.  We have chosen to use
$\alpha = 0.25$ and $\beta = 1.0$ as our standard
values.

TREESPH contains
the option to use a version of the usual SPH
artificial viscosity that reduces the amount of artificial
viscosity in the presence of curl
(Balsara 1989; Balsara, et al.\ 1989; Benz 1990).
 Tests carried out by
Centrella \& McMillan (1993) using TREESPH to calculate
the gravitational radiation from the head-on collision
of identical polytropes showed that the best results
were obtained using this modified artificial viscosity.
We have therefore chosen to use it here.

In these simulations shocks occur in the outer regions of
the model as the bar and spiral arms develop.
The effects of changing the artificial viscosity
coefficients can be seen by comparing models T3 and T4.
 Both of these runs have
$N = 15,648$ and started from the same initial equilibrium
model; see Table~\ref{T_models_params}.  Run T3 has
our standard values
$\alpha = 0.25$ and $\beta = 1.0$, while T4 uses the larger
values $\alpha = 1.0$ and $\beta = 2.0$. Models
T3 and T4 are very similar in visual appearance and
in their bulk properties, having the
same core values in the final state; see
Table~\ref{T-Models_data-hydro}.  The behavior of
the Fourier components of the density
 in both models is also similar.
Although the bar
mode growth rate is slightly lower in T4, this may not
be significant due to the element of subjectivity involved
in calculating it.
 As expected,
the energy conservation in T4 is improved due to the
greater smoothing of shocks by the larger artificial
viscosity.  In addition, T4 requires significantly
more CPU time than T3, due to the stability
requirement on the particle timesteps in the presence
of artificial viscosity (Hernquist \& Katz 1989).

The gravitational waveforms and luminosities for runs T3 and
T4 are similar, except that the quantities in T4 have lower
amplitudes than in T3.   This can be compared with
the results of Zhuge, Centrella, \& McMillan (1994), who
used TREESPH to simulate binary neutron star coalescence.
In their models, the stars merge and coalesce into a rotating
bar-like structure. Spiral arms form as mass is shed from the
ends of the bar; the arms then expand and merge into a disk
around the central object.  They found that, during this
spiral arm stage, the amplitude of the gravitational
waveforms decreases as the amount of artificial viscosity
is increased.

For comparison, run T5 also has the same number of particles
and began with the same initial state as T3, but
has no explicit artificial
viscosity, with $\alpha = \beta = 0$.  In this case,
the kinetic energy of the particles passing through
shocks is not correctly converted into heat, resulting
in large post-shock oscillations.
Table~\ref{T_models_params} shows that the energy conservation
errors are more than twice as large as those for run T3.
 The growth of the Fourier components is
similar, but the spiral arms spread out and
 merge more quickly after
the bar mode peaks in T5 compared to T3.
The final core $\varpi \le R_{\rm eq}$
has less mass and angular momentum than in run T3.
The values of the stability parameter $\tau$ for both
the core and the entire system are about the same for
T5 and T3.  However, T5 has a much larger kinetic energy
due to the large post-shock oscillations.  This gives
$T_{\rm total}/|W| = 0.31$ for the whole system compared
with $T_{\rm total}/|W| = 0.26$
for T3. The waveforms and
luminosity profiles for model T5 are similiar to T3 until
around the time that the amplitude of the bar mode reaches
its peak value; afterwards,
 the profiles for T5 become more noisy.

Finally, we compare models T6 and T3 to see the
effects of starting with a different
initial model.  Run T6 uses
the same values of $\alpha$ and $\beta$ as T3
and $N = 16,000$ particles; this is the model that
was presented in Houser, Centrella, \& Smith (1994).
The initial conditions for this run were produced using
the random or rejection method.  As mentioned in
\S~\ref{init-cond}, this technique produces relatively
large density fluctuations about the equilibrium solution.
Thus the particles in the initial model are acted on by
forces that can have large deviations from their
expected values, which can lead to violent motions
(Lucy 1977).  Although
TREESPH does perform some smoothing of
the initial data (Hernquist \& Katz 1989),
significant fluctuations remain. This results in energy
conservation errors $\sim 4.3\%$, which are more than twice
as large as those in T3.
The amplitude of the bar mode in T6 starts growing from a
larger initial value and reaches a much broader
peak before dropping to a local minimum again.  In
particular, there is a relatively short time interval during
which we can define a linear growth region for
$\ln |C|$ so we must
use caution in interpreting the
relatively large growth rate
reported in Table~\ref{T-Models_data-hydro}.
Run T6 also has more mass and angular momentum in the
core at the final state than T3,
although the stability parameter has the same value in
both cases.
Comparison of the gravitational wave data shows that
the gravitational wave burst begins immediately in
T6, compared to a starting time $\sim 10 t_{\rm D}$
for T3. Since  T6 goes out to
$26.9 t_{\rm D}$ whereas T3 goes out to $35 t_{\rm D}$,
the final states are roughly equivalent and can be
compared meaningfully.
  The gravitational wave quantities
for T6 do show the signature of a rotating nonaxisymmetric
core after the burst, although there is not as much
detail as in T3.  The amplitudes of the waveforms and
the luminosities are both lower for T6.
See Table~\ref{T-Models_data-GW}.

It is interesting to compare our results with those of
DGTB, who evolved the dynamical bar
instability using an SPH code with $N=2000$ particles
and smoothing lengths that are allowed to vary in time
but are the same for all particles.
Plots of the particle
positions projected onto the $x-y$ plane
for the case $\tau \approx 0.33$ are visually similar to
our Run T1.  They report that the system has
$\tau = 0.247$ at the end of their run, which is
essentially the same as our result for T1.

In summary, all the models run with
TREESPH show the development of the bar and the
spiral arm pattern.
 Models T1 - T4 and T7 conserve total
energy to $\lesssim 2\%$, with T5 and T6 conserving energy
to $\sim 4\%$.  In all cases, angular momentum is conserved
to $\lesssim 0.1\%$.
    The bar mode growth rates for models
T2 - T7 are the same to within $\sim 6\%$; we attribute
these differences largely to the element of
subjectivity inherent in choosing the region of linear
growth for $d \ln |C|/dt$.  The anomalous growth rate
found for T1 is due to the lack of a clearly defined
linear growth region.  The properties of the final
cores $\varpi \le R_{\rm eq}$ in the models are
remarkably similar, especially if runs T5 (with no
artificial viscosity) and T6 (with a noisy initial state)
are excluded.  Finally, the amplitudes of the
gravitational wave quantities increase as $N$ increases;
higher resolution runs, or models with non-equal-mass
particles are needed to achieve convergence in these
quantities.

\section{Comparison of Eulerian and SPH Results}
\label{comparisons}

We turn now to a comparison between the results of the
Eulerian and SPH codes.  To accomplish this we focus on
our standard
models E2 and T7.  There is no simple measure for comparing
the resolution of these two models since the underlying
fluid descriptions are so different.  In run E2, the fluid
initially occupies $\sim 14,400$ zones; after flowing
through the grid during the development of the bar mode,
the fluid occupies $\sim 66,200$ zones at the end of the
run.  In contrast, the fluid in run T7 is discretized
into $N = 32,914$ equal-mass particles.  As the system
evolves, these particles move through space, with the
smoothing length of each particle continually adjusted
to keep the number of nearest neighbors approximately
constant.  For our purposes we simply note that the
number of particles used in T7 is comfortably between
the initial and final number of zones occupied by the
fluid in E2, and proceed with the comparison.

Examination of Tables~\ref{E_models_params}
and~\ref{T_models_params} shows that T7 conserves both
total energy and angular momentum better
than does E2.  Run T7 also uses a larger amount of
CPU time.  However, as demonstrated in
\S~\ref{sec-eul-vary}, the energy and angular momentum
conservation for E2 both improve when relaxed vacuum
conditions are used, although at the expense of a
larger usage of CPU time.

In both models, the growth of the $m=2$ mode produces a rotating
bar that develops spiral arms as mass is shed from the ends
of the bar. The bar and spiral arms exert gravitational torques
that cause angular momentum to be transported outward.
The spiral arms expand supersonically and merge together,
causing shock heating and dissipation in the disk surrounding
the central core.  The system remains highly flattened and
evolves toward a {\em nearly} axisymmetric final state in
both models.

The density contours for E2 in Figure~\ref{E-contours}
and for T7 in Figure~\ref{T7-contours} can be
compared directly, since the contours for T7 were made
by interpolating the particle model onto the grid used
for E2 and these figures use the same contour levels.
The most striking visual difference lies in the fact that
T7 expands much more than does E2.
 These differences in
the amount of expansion account
 for the differences in
the core ($\varpi \le R_{\rm eq}$)
masses and angular momenta given in
Tables~\ref{E-Models_data-hydro}
and~\ref{T-Models_data-hydro}.
When the net transport of angular momentum is considered,
the behavior of the models is more similar.
At the final time, for example, in model
E2 $90\%$ of the mass has $73\%$ of the angular
momentum and in T7 $90 \%$ of the mass has
$71 \%$ of the angular momentum.

The growth of the bar mode is shown quantitatively
in Figure~\ref{E2-bar}
for run E2 and in Figure~\ref{T7-bar} for run T7.  In both
cases there is a relatively long period of linear growth
for the bar mode amplitude $\ln |C|$. The measured values
in this linear growth region are
$d \ln|C|/dt = 0.58 t_{\rm D}^{-1}$ for E2 and
$d \ln|C|/dt = 0.51 t_{\rm D}^{-1}$ for T7. The
larger slope for E2 is closer to the analytic TVE
value $d\ln|C|/dt = 0.728 \pm 0.038 t_{\rm D}^{-1}$
given by TDM.  As was discussed in \S~\ref{sec-eul-vary},
numerical diffusion can cause the growth rate given by
a simulation to be lower than the expected
value (TDM).  This suggests that the SPH code has
more numerical diffusion than the Eulerian code;
further studies of this would be very useful.
The eigenfrequency
 is $\sigma_2 = 1.8 t_{\rm D}^{-1}$ for E2
and $\sigma_2 = 1.9 t_{\rm D}^{-1}$ for T7; both
of these values are within the range
 $\sigma_2 = 1.892 \pm 0.094 t_{\rm D}^{-1}$
given for the TVE result (TDM).

In both runs, the bar mode amplitudes reach
their maximum values, drop off to local minima, and
then grow again.  The bar mode in T7 reaches a higher
value than in E2, but the peak is broader.  In run E2
the amplitude reaches a second peak at a lower amplitude
than the first, and then drops again.
In both cases, the $m=4$ mode starts growing after the
bar mode, grows at a faster exponential rate, and then
peaks at about the same time as the bar mode before
dropping off;
cf. Figures~\ref{E2-modes}(c)
and~\ref{T7-modes}(c). For E2 we find
$d \ln (|C_4|/|C_0|)/dt = 1.1 t_{\rm D}^{-1}$
and $\sigma_4 = 3.4 t_{\rm D}^{-1}$ and
for T7 we get
$d \ln (|C_4|/|C_0|)/dt = 1.1 t_{\rm D}^{-1}$
and $\sigma_4 = 3.8 t_{\rm D}^{-1}$.
In both runs we find that the pattern speeds for
these two modes are about the same, indicating
that the $m=4$ mode is a harmonic of the $m=2$ mode.

One interesting difference between the models can be seen by
comparing the behavior of the $m=1$ and $m=3$ Fourier components
of the density.  Figure~\ref{E2-modes} shows that these
disturbances grow in run E2 whereas Figure~\ref{T7-modes} shows
that they do not grow in run T7.  We do not have an explanation
for this behavior. However, given the importance of the $m=1$
mode in recent work (Bonnell 1994; Bonnell \& Bate 1994; PDD),
this question deserves further study

The overall behavior of the stability parameter $\tau$
and $T_{\rm total}/|W|$ is similar in both E2 and T7, as
can be seen by comparing Figures~\ref{tau-vs-t-E2}
and~\ref{tau-vs-t-T7}. In both cases,
the final value of the stability parameter
is in the range
$\tau_{\rm s} < \tau < \tau_{\rm d}$.
We therefore expect that the system will develop
a secular bar instability when the effects of gravitational
radiation reaction are included in the hydrodynamical equations
(cf.\ Lai \& Shapiro 1995).

The gravitational radiation in both models is dominated
by a strong feature that corresponds to the initial growth
of the bar mode, with the peak amplitudes of both the waveforms
and luminosity being higher in E2 than in T7.  The radiation
emitted after the initial burst shows a secondary feature
in both models.  In E2 the radiation has
a double-burst structure; this behavior is less distinct in
T7, although the secondary feature is present.
We showed in \S~\ref{sec-eul-vary}
that the gravitational radiation amplitudes for E2
decrease somewhat
if we halve the number of angular zones or change
the treatment of the vacuum zones
to allow the model to expand more.  Also, \S~\ref{sec-sph-vary}
showed that the gravitational wave
 amplitudes for the SPH models increase
and approach the E2 amplitude as
the particle number increases, and that they have not
yet converged for this set of runs.
 Further work with both codes, including higher resolution
runs, is needed to resolve these issues

\section{Conclusions}
\label{conclusions}

We have carried out 3-D numerical simulations of the
dynamical bar instability  in a rapidly rotating star
and the resulting gravitational radiation using both an
Eulerian finite-difference code with a cylindrical grid
and monotonic advection, and an SPH code with variable
smoothing lengths and a hierarchical tree method for
calculating the gravitational acceleration.  The star
is initially modeled as a polytrope with index $n=3/2$
and $\tau \approx 0.30$.  In both codes the gravitational
field is purely Newtonian and the gravitational radiation
is calculated using the quadrupole formula.  The back reaction
of the gravitational radiation on the fluid is not included.

In both codes the dynamical instability of the $m=2$ mode produces
a rotating bar-like structure. Spiral arms develop as mass is
shed from the ends of the bar, and gravitational torques
cause angular momentum to be transported outward.  The
spiral arms expand supersonically and merge, causing shock
heating in the outer regions.  At the end of the simulations,
both codes agree that
the system consists of a {\em nearly} axisymmetric central core
surrounded by an extended disk, and has $\tau \sim 0.25$.

It is interesting to compare our results with those from the
earlier study by DGTB that considered rapidly rotating
$n=3/2$ polytropes with $\tau \approx 0.33$ and
$\tau \approx 0.38$.  The Eulerian codes used in that work
were quite diffusive. They also did not solve
an energy equation and thus had no means of handling
self-consistently the shocks that form in the outer
regions.  Also, a certain amount of mass was
allowed to leave the grid.
In our Eulerian runs, shocks are
handled using an artificial viscosity and no mass
is allowed to leave the grid.
Our main difficulty in the Eulerian code is with the expansion
of the model into the vacuum.  We believe this can be alleviated
with the use of better vacuum conditions.

The SPH code used by DGTB had a smoothing length
that was the same for all particles.  Their runs were
also limited to a fairly small number of particles.
The use of variable smoothing lengths and the hierarchical
tree method for calculating the gravitational accelerations
in TREESPH allows us to evolve more particles.
All the SPH models have no grid and
expand freely.  However, the fluid description can break down
in the low density outer regions due to lack of resolution.

The simulations of DGTB showed the development
of the bar instability with spiral arms and transport of
angular momentum.  However, their codes differed in the
final outcome of the models in the low density outer regions.
In their longest Eulerian run, most of the low density material
formed a fairly narrow ring around the central remnant.
(This outcome was also seen by Williams \& Tohline (1988)
for polytropes having $n=0.8$ and $n=1.8$ with $\tau = 0.31$.)
However, in their SPH runs the low density material formed an
extended disk.

We are not sure what causes these differences in the
final outcome.
Further study, perhaps including longer runs with our
codes, is needed to resolve these issues.

Interestingly, DGTB concluded that their SPH runs had less
numerical diffusion than their Eulerian simulations.  We
believe this is due to the low-order differencing in their
Eulerian codes.
In our study, the use of
higher-order differencing and monotonic advection
in the Eulerian code resulted in
much less numerical diffusion than seen in the earlier studies.
(See also PDD.)
Comparison of the bar growth rates for our standard Eulerian
and SPH runs suggests that
 the SPH code has more numerical diffusion.

We agree with DGTB that the SPH code is
 easier to implement and use.  We find that
the bulk properties of the model can be obtained
at very low cost using $N \sim 2000$ particles,
although more particles are needed for a good
measure of the bar mode growth rate.
We did not carry out any very low resolution
studies with our Eulerian code.  However, we find
that the peak amplitudes of the gravitational radiation
quantities in both the Eulerian and SPH
codes increase as the resolution
is increased.  Overall, for comparable resolution,
the cost of the Eulerian and SPH runs is
similar.

We note that although the version of SPH
used here allows the particle smoothing lengths to
vary in both space and time, the terms describing these
changing smoothing lengths are not explicitly incorporated
into the dynamical equations (Hernquist \& Katz 1989).
Although this situation is typical of most SPH codes
currently used in astrophysics, it does present a potentially
serious deficiency in the method.  Recent work to incorporate
these terms into the dynamical equations
(e.g.\ Nelson \& Papaloizou 1995)
may lead to substantial improvements in the SPH method.

Finally, the suitability of a particular
numerical method must be determined in the context of
the astrophysical system being modeled, as well as
the resources available to the investigators.
We hope that this study, along with related work
comparing the results of Eulerian and SPH
hydrodynamic codes in modeling stellar collisions
(Davies, et al. 1993) and the growth of
structure in cosmology (Kang, et al. 1994),
will help others to find the best approach
for their applications.

\acknowledgments

We thank S. Clancy, S. Cranmer and S. McMillan
for helpful conversations, and L. Hernquist for supplying
a copy of TREESPH.  We are pleased to acknowledge the
helpful comments of the referee R. Durisen, which contributed
to improving this paper. This work was supported in
part by NSF grant PHY-9208914.
The simulations were
carried out on the Cray C90 at Pittsburgh
Supercomputing Center under grant
PHY910018P.
\newpage

\noindent
Postal Addresses:

\noindent
Scott C. Smith: Dept. of Physics, LaSalle University, Philadelphia,
PA  19141

\noindent
Janet L. Houser and Joan M. Centrella: Dept. of Physics and
Atmospheric Science, Drexel University, Philadelphia, PA  19104

\newpage
\clearpage

\renewcommand{\arraystretch}{1.5}

\begin{table}[p]
\begin{center}
\begin{tabular}{ccccccccccc}
\tableline
Run & $N_{\varpi}$ & $N_z$ & $N_{\varphi}$
    & $N_{\rm star,i}$ & $N_{\rm star, f}$ &
    $\tau_{\rm i}$ &
time &
$ \left | \frac{E_{\rm i} - E_{\rm f}}{E_{\rm i}} \right | $
&
$ \left | \frac{J_{\rm i} - J_{\rm f}}{J_{\rm i}} \right | $
& CPU
\\
\tableline
E1 & 64 & 32 & 32  & 7200 & 24,600 &
   0.30 & $34  t_{\rm D}$ & .035 & .072 &
   8.2 hr  \\
E2 & 64 & 32 & 64  & 14,400 &
66,200 & 0.30 & $34 t_{\rm D}$ & .043 & .086 &
   29.7 hr\\
${\rm E}2^{\prime}$ & -- & -- & --  & --- &
41,500 & -- & $23.9 t_{\rm D}$ & .020 & .049 &
   20.5 hr \\
E3 & 64 & 32 & 64  & 14,400 &
110,600 & 0.30  & $23.9 t_{\rm D}$ & .015 & .034 &
   29.5 hr \\
\tableline
\end{tabular}
\end{center}
\caption{Properties of the Eulerian models. $N_{\varpi}$,
$N_z$, and $N_{\varphi}$ are the number of grid zones in the
$\varpi$, $z$, and $\varphi$ directions, respectively.
$N_{\rm star}$ is the approximate number of zones occupied
by the matter distribution.  The subscripts ``i'' and
``f'' denote the initial and final states of the model.
 The stability parameter at the
initial time, calculated on the Eulerian grid, is
$\tau_{\rm i}$. The
duration of the run is measured in units of the
dynamical time $t_{\rm D}$ and listed in the column
labeled ``time''.  All models were run on a Cray C90;
the amount of CPU time used is given for the duration
of the run. The quantities given in the row
labeled ${\rm E}2^{\prime}$ are the values for model E2 at
the intermediate time $23.9 t_{\rm D}$.  Model E3 is the
same as E2 except for the use of relaxed constraints in
the vacuum conditions.
}
\label{E_models_params}
\end{table}

\begin{table}[p]
\begin{center}
\begin{tabular}{ccccccc}
\tableline
Run & $d \ln |C|/dt$ & $\sigma_2$ &
 $M_{\rm core, f}$ &
 $J_{\rm core, f}$  & $\tau_{\rm core, f}$
&  $\tau_{\rm f}$ \\
 \tableline
E1 & 0.53  & 1.7 & 94\% & 78\%  & 0.18 & 0.19 \\
E2 & 0.58  & 1.8 & 96\% & 86\% & 0.24 & 0.24 \\
${\rm E}2^{\prime}$ & --  & --  &
96\% & 87\% & 0.25 & 0.26 \\
E3 & 0.57 & 1.8 & 95\% & 83\% & 0.25 & 0.26  \\
\tableline
\end{tabular}
\end{center}
\caption{Hydrodynamical and bar mode results for
the Eulerian models.  The quantities in the row labeled
${\rm E}2^{\prime}$ are the values
for run E2 at the intermediate
time $23.9 t_{\rm D}$; cf.\
Table~\protect{\ref{E_models_params}}. The bar growth rate
$d \ln |C|/dt$ and the eigenfrequency $\sigma_2$ are
calculated for the ring $\varpi = 0.362 R_{\rm eq}$ in
zone $j = 10$ in the equatorial plane; both are
in units of $t_{\rm D}^{-1}$.
 The core refers to matter within
cylindrical radius $\varpi = R_{\rm eq}$,
where $R_{\rm eq}$ is the equatorial radius of the initial
model, and the subscript ``f''
denotes the final state of the model.
Note that $J_{\rm core,f}$ is normalized using the value of the
total angular momentum of the system at the final time, which
is less than the initial value due to non-conservation;
see  Table~\protect{\ref{E_models_params}}.
}
\label{E-Models_data-hydro}
\end{table}

\begin{table}[p]
\begin{center}
\begin{tabular}{cccccc}
\tableline
Run & max $|rh|$ & max $L$ & $(\Delta E/M)_{\rm f}$
 &  max $dJ_z/dt$ & $(\Delta J/J_0)_{\rm f}$ \\
 \tableline
E1 & 0.63 & 0.20 & 0.68 & 0.18 & 1.6  \\
E2 & 0.68 & 0.21 & 0.93 & 0.20 & 2.3  \\
${\rm E}2^{\prime}$ & -- & -- & 0.77 & -- & 2.0 \\
E3 & 0.62 & 0.17 & 0.63 & 0.16 & 1.6 \\
\tableline
\end{tabular}
\end{center}
\caption{Gravitational wave results for
the Eulerian models. We use $c = G = 1$.
The quantities in the row labeled
${\rm E}2^{\prime}$ are the values
for run E2 at the intermediate
time $23.9 t_{\rm D}$; cf.\
Table~\protect{\ref{E_models_params}}.
The values of the
peak gravitational wave amplitudes $|rh|$ are in units of
$M^2/R_{\rm eq}$.  The values of the maximum luminosity
$L$ are in units of $(M/R_{\rm eq})^5$, and the values of the
total energy emitted during the duration of the run
$(\Delta E/M)_{\rm f}$ are in units of $(M/R_{\rm eq})^{7/2}$.
The values of the
maximum $dJ_z/dt$ are in units of $M (M/R_{\rm eq})^{7/2}$.
The quantity $(\Delta J_z/J_0)_{\rm f}$ is the total
angular momentum emitted as gravitational radiation
 normalized by the initial total angular
momentum, and has units of $(M/R_{\rm eq})^{5/2}$.
}
\label{E-Models_data-GW}
\end{table}

\begin{table}[p]
\begin{center}
\begin{tabular}{cccccccccc}
\tableline
Run & $N$ & type & $\alpha$ &  $\beta$ &
    $\tau_{\rm i}$  & time &
$ \left | \frac{E_{\rm i} - E_{\rm f}}{E_{\rm i}} \right | $
&
$ \left | \frac{J_{\rm i} - J_{\rm f}}{J_{\rm i}} \right | $
     & CPU \\
\tableline
T1 & 2061 & cold & 0.25 & 1.0 & 0.305
& $35 t_{\rm D}$ & 0.022 & $\lesssim .001$ &
   1.01 hr \\
T2 & 8728 & cold & 0.25 & 1.0 & 0.308
& $35 t_{\rm D}$ & 0.018 & $\lesssim .001$ &
   8.65 hr \\
T3 & 15,648 & cold & 0.25 & 1.0 & 0.314
& $35 t_{\rm D}$ & 0.018 & $\lesssim .001$ &
   16.9 hr \\
T4 & 15,648 & cold & 1.0 & 2.0 & 0.314
& $35 t_{\rm D}$ & 0.011 & $\lesssim .001$ &
   23.0 hr \\
T5 & 15,648 & cold & 0 & 0 & 0.314
& $35 t_{\rm D}$ & 0.043 & $\lesssim .001$ &
   16.7 hr \\
T6 & 16,000 & random & 0.25 & 1.0 & 0.299
 & $26.9 t_{\rm D}$ &
   0.043 &  $\lesssim .001$ & 11.6 hr  \\
T7 & 32,914 & cold & 0.25 & 1.0 & 0.316
& $35 t_{\rm D}$ & 0.018 &  $\lesssim .001$ &
   44.6 hr \\
\tableline
\end{tabular}
\end{center}
\caption{Properties of the SPH models.  All models used the
``cold'' initial conditions except T6; the initial conditions
for this model were produced using the random method.  $N$ is
the number of particles in the model, and $\alpha$ and $\beta$
are, respectively, the coefficients of the linear and quadratic
terms in the artificial viscosity.
The other quantities are defined as in
Table~\protect{\ref{E_models_params}}.
}
\label{T_models_params}
\end{table}

\begin{table}[p]
\begin{center}
\begin{tabular}{ccccccc}
\tableline
Run & $d \ln |C|/dt$ & $\sigma_2$ &
$M_{\rm core, f}$ &
 $J_{\rm core, f}$   & $\tau_{\rm core, f}$
& $\tau_{\rm f}$ \\
 \tableline
T1 & 0.66 & 2.1 & $91\%$ & $73\%$ & 0.24 & 0.25 \\
T2 & 0.51 & 1.9 & $92\%$ & $74\%$ & 0.24 & 0.26 \\
T3 & 0.53 & 1.9 & $92\%$ & $73\%$ & 0.25 & 0.26 \\
T4 & 0.51 & 1.9 & $92\%$ & $73\%$ & 0.25 & 0.26 \\
T5 & 0.54 & 1.9 & $89\%$ & $67\%$ & 0.24 & 0.26 \\
T6 & 0.54 & 1.8 & $95\%$ & $81\%$ & 0.25 & 0.26 \\
T7 & 0.51 & 1.9 & $90\%$ & $72\%$ & 0.25 & 0.26 \\
\tableline
\end{tabular}
\end{center}
\caption{Hydrodynamical and bar mode results for the SPH
models. The quantities are defined as in
Table~\protect{\ref{E-Models_data-hydro}}.
The large growth rate for T1 is anomalous; see the
text for details.
}
\label{T-Models_data-hydro}
\end{table}

\begin{table}[p]
\begin{center}
\begin{tabular}{cccccc}
\tableline
Run & max $|rh|$ & max $L$ & $(\Delta E/M)_{\rm f}$
 &  max $dJ_z/dt$ & $(\Delta J_z/J_0)_{\rm f}$ \\
 \tableline
T1 & 0.26 & 0.030 & 0.21 & 0.015 & 0.24  \\
T2 & 0.43 & 0.078 & 0.46 & 0.039 & 0.54  \\
T3 & 0.47 & 0.091 & 0.55 & 0.045 & 0.66 \\
T4 & 0.41 & 0.066 & 0.39 & 0.035 & 0.48  \\
T5 & 0.48 & 0.10  & 0.52 & 0.050 & 0.59  \\
T6 & 0.38 & 0.051 & 0.31 & 0.027 & 0.42 \\
T7 & 0.53 & 0.12  & 0.87 & 0.060 & 1.0 \\
\tableline
\end{tabular}
\end{center}
\caption{Gravitational wave results for the SPH
models.
The quantities are defined as in
Table~\protect{\ref{E-Models_data-GW}}.
}
\label{T-Models_data-GW}
\end{table}

\clearpage
\newpage

\begin{figure}[p]
\caption{Density contours in the equatorial plane for model E2.
The contour levels are the same in all frames.
The contours are spaced a factor of 10 apart, and go down to
3 decades below the maximum (central) density at the initial
time $t = 0$.  Since the maximum density increases slightly
during the run, there are 4 contours shown for the later
times, with the innermost contour being at the initial
maximum density.  The model rotates in the counterclockwise
direction.  The circular boundary of the plotted region
is set at $\varpi = 2.5 R_{\rm eq}$.
}
\label{E-contours}
\end{figure}

\begin{figure}[p]
\caption{The mass fraction $m(\varpi)$ is shown for
model E2 at the initial time (dot-dashed line), the
 intermediate time
 $t = 21.1 t_{\rm D}$ (dashed line),
and the final time $t = 34t_{\rm D}$ (solid line).
The total mass is $M$.
}
\label{E2-m}
\end{figure}

\begin{figure}[p]
\caption{The angular momentum $J(\varpi)$ is
shown for the model E2 at the initial time (dot-dashed line),
the intermediate time
$t = 21.1 t_{\rm D}$ (dashed line),
and the final time $t = 34t_{\rm D}$ (solid line).
Note that $J(\varpi)$ is normalized by the total
angular momentum in the system at the time.
}
\label{E2-j}
\end{figure}

\begin{figure}[p]
\caption{The growth of the bar mode for model E2.
The bar amplitude $\ln |C|$ is shown versus time for
the ring $\varpi = 0.362 R_{\rm eq}$ in zone $j=10$ in the
equatorial plane $z=0$.  The growth rate in the
region where $\ln |C|$ is growing linearly with time is
$d \ln|C|/dt = 0.58 t_{\rm D}^{-1}$.
}
\label{E2-bar}
\end{figure}

\begin{figure}[p]
\caption{The growth of various Fourier components
of the density for model E2.
The amplitudes have been calculated for the same ring
used in Figure~\protect{\ref{E2-bar}}, and are shown
versus time.  In each case, the bar mode amplitude
is plotted as a solid line for comparison.
(a) $m = 1$ (b) $m = 3$ (c) $m = 4$
}
\label{E2-modes}
\end{figure}

\begin{figure}[p]
\caption{The behavior of $T/|W|$ is shown as a function
of time for model E2.  The solid line shows the stability
parameter $\tau = T_{\rm rot}/|W|$ and
the dashed line shows $T_{\rm total}/|W|$.
}
\label{tau-vs-t-E2}
\end{figure}

\begin{figure}[p]
\caption{Gravitational waveforms
 for an observer on the axis at
$\theta = 0$ and $\varphi = 0$ for model E2.
We use $c = G = 1$.
(a) $r h_+$ (b) $r h_{\times}$
}
\label{E2-waves}
\end{figure}

\clearpage

\begin{figure}[p]
\caption{Various gravitational wave quantities are shown
for run E2. We use $c = G = 1$.
 (a) Gravitational wave luminosity $L$.
(b) The energy $\Delta E/M$ emitted as gravitational
radiation.  (c) The rate $dJ_z/dt$ at which angular
momentum is carried away by the waves.  (d) The
angular momentum $\Delta J/J_0$ lost to gravitational
radiation. Here, $J_0$ is the initial total angular
momentum.
}
\label{E2-lum-E}
\end{figure}

\begin{figure}[p]
\caption{Density contours in the equatorial plane at
$t = 23.9 t_{\rm D}$. The contour levels are the same
as in Figure~\protect{\ref{E-contours}}.
 (a) Model E2  (b) Model E3
(with relaxed vacuum conditions)
}
\label{contours-E2-E3}
\end{figure}

\begin{figure}[p]
\caption{Particle positions are shown projected onto the
equatorial plane for various times in the evolution of
model T7. The vertical axis is
$y/R_{\rm eq}$ and the horizontal axis is
$x/R_{\rm eq}$.
 The system rotates in the
counterclockwise direction.
}
\label{T7-dot-plots}
\end{figure}

\begin{figure}[p]
\caption{Particle positions are shown projected onto the
$x - z$ plane at the final time $t = 35 t_{\rm D}$ of
model T7.  Shock heating during the spiral arm expansion
has caused the disk to puff up.
}
\label{T7-meridional}
\end{figure}

\begin{figure}[p]
\caption{Density contours in the equatorial plane
for model T7.  The density has been interpolated
 onto the cylindrical grid
used for model E2, and the contour levels are the same as those
used in Figure~\protect{\ref{E-contours}}.
}
\label{T7-contours}
\end{figure}

\begin{figure}[p]
\caption{The mass fraction $m(\varpi)$ is shown for
model T7 at the initial time (dot-dashed line), the
intermediate time
 $t = 18.2 t_{\rm D}$ (dashed line),
and the final time $t = 35t_{\rm D}$ (solid line).
Only the region $\varpi \le 5R_{\rm eq}$ is plotted.
The total mass is $M$.
}
\label{T7-m}
\end{figure}

\begin{figure}[p]
\caption{The angular momentum $J(\varpi)$ is
shown for the model T7 at the initial time (dot-dashed line),
intermediate time
$t = 18.2 t_{\rm D}$ (dashed line),
and the final time $t = 35t_{\rm D}$ (solid line).
Only the region $\varpi \le 5R_{\rm eq}$ is plotted.
The total initial angular momentum is $J_0$.
}
\label{T7-j}
\end{figure}

\begin{figure}[p]
\caption{The growth of the bar mode for model T7.  The
particle model is interpolated onto the cylindrical grid
used for model E2, and the bar amplitude is calculated
as in Figure~\protect{\ref{E2-bar}}.
The growth rate in the
region where $\ln |C|$ is growing linearly with time is
$d \ln|C|/dt = 0.51 t_{\rm D}^{-1}$.
}
\label{T7-bar}
\end{figure}

\begin{figure}[p]
\caption{The growth of various Fourier components of
the density for model T7.
In each case, the bar mode amplitude
is plotted as a solid line for comparison.
(a) $m = 1$ (b) $m = 3$ (c) $m = 4$
}
\label{T7-modes}
\end{figure}

\begin{figure}[p]
\caption{The behavior of $T/|W|$ is shown as a function
of time for model T7.  The solid line shows the stability
parameter $\tau = T_{\rm rot}/|W|$ and
the dashed line shows $T_{\rm total}/|W|$. The initial
oscillations in $T/|W|$ are
caused by adjustments in the model due to residual noise
in the initial conditions.
}
\label{tau-vs-t-T7}
\end{figure}

\begin{figure}[p]
\caption{Gravitational waveforms
 for an observer on the axis at
$\theta = 0$ and $\varphi = 0$ for model T7.
We use $c = G = 1$.
(a) $r h_+$ (b) $r h_{\times}$
}
\label{T7-waves}
\end{figure}

\begin{figure}[p]
\caption{Various gravitational wave quantities are shown
for run T7. We use $c = G = 1$.
 (a) Gravitational wave luminosity $L$.
(b) The energy $\Delta E/M$ emitted as gravitational
radiation.  (c) The rate $dJ_z/dt$ at which angular
momentum is carried away by the waves.  (d) The
angular momentum $\Delta J/J_0$ lost to gravitational
radiation.
}
\label{T7-lum-E}
\end{figure}

\end{document}